\documentclass[preprint,12pt,prb]{revtex4}%
\usepackage{amsfonts}
\usepackage{amsmath}
\usepackage{amssymb}
\usepackage{graphicx}%
\providecommand{\U}[1]{\protect\rule{.1in}{.1in}}
\newtheorem{theorem}{Theorem}

\begin{document}
\preprint{UATP/1101}
\title{A Rigorous Derivation of the Entropy Bound and the Nature of Entropy Variation
for Non-equilibrium Systems during Cooling}
\author{P. D. Gujrati}
\email{pdg@uakron.edu}
\affiliation{Department of Physics, Department of Polymer Science, The University of Akron,
Akron, OH 44325}

\begin{abstract}
We use rigorous non-equilibrium thermodynamic arguments to prove (i) the
residual entropy of any system is bounded below by the experimentally
(calorimetrically) determined absolute temperature entropy, which itself is
bounded below by the entropy of the corresponding equilibrium (metastable
supercooled liquid) state, and (ii) the instantaneous entropy cannot drop
below that of the equilibrium state. The theorems follow from the second law
and the existence of internal equilibrium and refer to the thermodynamic
entropy. They go beyond the calorimetric observations by Johari and Khouri [J.
Chem. Phys. \textbf{134}, 034515 (2011)] and others by extending them to all
non-equilibrium systems regardless of how far they are from their equilibrium
states. We also discuss the statistical interpretation of the thermodynamic
entropy and show that the conventional Gibbs or Boltzmann interpretation gives
the correct thermodynamic entropy even for a single sample regardless of the
duration of measurements.

\end{abstract}
\date[January 24, 2011]{}
\maketitle

\section{Introduction}

In a recent publication, Johari and Khouri \cite{Johari} used upper and lower
bounds on the isobaric entropy $S(T_{0})$ of a glass as a function of the
temperature $T_{0}$ of the surrounding medium, obtained by using the measured
heat capacity during cooling and heating to argue for the reality of the
residual entropy $S_{\text{R}}$, as the latter has become a highly debated
issue in the
literature.\cite{Note1,Jackel,Palmer,Hemmen,Thirumalai,Kivelson,Gupta,Goldstein,Gupta1,Mauro}
For a brief review of the history of the residual entropy and the current
controversy, we refer the reader to Goldstein,\cite{Goldstein} Gutzow and
Schmelzer,\cite{Gutzow-Schmelzer} Nemilov,\cite{Nemilov} and the recent
reviews\cite{GujratiResidualentropy,Gujrati-Symmetry} by us; see also below.
The existence of a non-zero residual entropy ($S_{\text{R}}>0$) is very common
in Nature, and does not violate Nernst's postulate, as the latter is
applicable only to equilibrium states with a \emph{non-degenerate} ground
state; see Sect. 64 in Landau and Lifshitz.\cite{Landau} Its existence was
first theoretically demonstrated by Pauling and Tolman;\cite{Pauling} see also
Tolman. \cite{Tolman} In addition, the existence of the residual entropy has
been demonstrated rigorously for a very general spin model by Chow and
Wu.\cite{Chow} The residual entropy for glycerol was observed by Gibson and
Giauque\cite{Giauque-Gibson} and for ice by Giauque and Ashley.\cite{Giauque}
Pauling \cite{Pauling-ice} provided the first numerical estimate for the
residual entropy for ice, which was later improved by Nagle.\cite{Nagle}
Nagle's numerical estimate has been recently verified by
simulation.\cite{Isakov,Berg} The numerical simulation carried out by Bowles
and Speedy\cite{Speedy} for glassy dimers also supports the existence of a
residual entropy. Richet \cite{Richet} uses the Adam-Gibbs theory to justify
the residual entropy. Thus, it appears that the support in favor of the
residual entropy is quite strong. We wish to emphasize that what is
customarily called the third law due to Nernst, according to which the entropy
must vanish at absolute zero, is merely a postulate and not a strict theorem
even in equilibrium.\cite{Landau,Gujrati-Nernst,Gujrati-Fluctuations} Indeed,
many exactly solved statistical mechanical models show a non-zero entropy at
absolute zero. However, as of yet, no experiment can be performed at absolute
zero to demonstrate the residual entropy; in all cases, some sort of
\emph{extrapolation} is required. This point should not be forgotten in the
following whenever we speak of measuring the residual entropy. In addition, we
will speak of the equilibrium state associated with a non-equilibrium state.
Depending on the context, the equilibrium state may represent a true
equilibrium state such as a crystal or a (time-independent) metastable state
such as the supercooled liquid. However, for the purpose of clarity, we will
consider the supercooled liquid in the following, but the arguments are
applicable to both cases.

\subsection{Controversy and Its Current Status\label{Sect_Controversy_Status}}

\begin{figure}
[ptb]
\begin{center}
\includegraphics[
trim=0.000000in -0.159333in -0.159305in 2.386659in,
height=2.8037in,
width=4.1788in
]%
{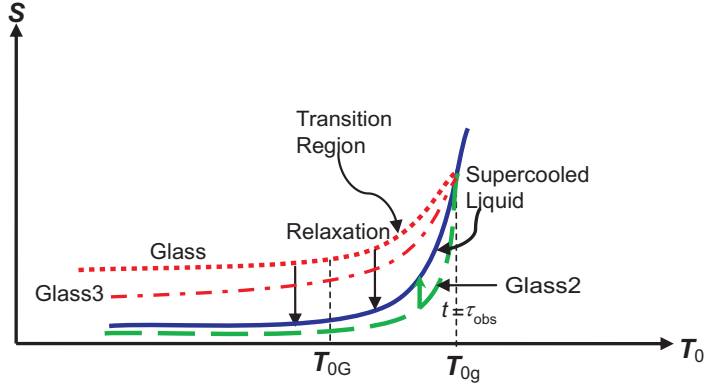}%
\caption{Schematic behavior of the entropy: equilibrated supercooled liquid
(solid curve) without any relaxation and three possible
glasses\ (Glass1-dotted curve, Glass2-dashed curve,\ Glass3-dash-dotted curve)
during vitrification as a function of the temperature $T_{0}$ of the medium.
Structures appear to freeze (over an extremely long period of time) at and
below $T_{0\text{G}}$; see text. The transition region between $T_{0\text{g}}$
and $T_{0\text{G}}$ over which the liquid turns into a glass has been
exaggerated to highlight the point that the glass transition is not a sharp
point.\ For $T_{0}<T_{0\text{g}}$, the non-equilibrium state undergoes
isothermal structural relaxation in time towards the supercooled liquid. For
Glass1, the two vertical downward arrows show isothermal structural relaxation
at two different temperatures, during which the entropy decreases. The same
will also happen for Glass 3. For Glass2, the entropy increases during
isothermal structural relaxation; see the upward arrow. The instantaneous
temperature $T(t)$ of the glass decreases towards the $T_{0}$ during
relaxation., so that the entropy is a function of $T(t)$ during relaxation.
The entropy of the supercooled liquid is shown to extrapolate to zero per our
assumption, but that of Glass1 to a non-zero value and of Glass2 to zero at
absolute zero. The entropy of Glass 3 may or may not vanish at absolute. The
possibility of an ideal glass transition, which does not affect our
conclusion, will result in a singular form of the solid curve. The second law
is used to support Glass1 and rule out Glass2; see the text. }%
\label{Fig_entropyglass}%
\end{center}
\end{figure}

It is surprising to see this controversy persist in the current literature
even though it seemed resolved a long time ago;\cite{Pauling,Tolman} we also
note somewhat recent attempts.\cite{Sethna,Sethna-Paper} The controversial
issue is the following: As the irreversibility does not allow for an exact
evaluation of the entropy, is it possible for the entropy to decrease by an
amount almost equal to $S_{\text{R}}$ within the glass transition region so
that the glass (see Glass2 in Fig. \ref{Fig_entropyglass}) would have a
vanishing entropy at absolute zero? The entropy for Glass2 follows from the
recent proposals by Kivelson and Reiss,\cite{Kivelson} and by Gupta and
Mauro;\cite{Gupta-JNCS,Gupta,Gupta1} see also Reiss.\cite{Reiss} Gupta and
Mauro\cite{Gupta-JNCS} conclude that
\begin{equation}
S\leq S_{\text{SCL}}\text{ \ \ \ } \label{GM_bound}%
\end{equation}
\ at and below the glass transition, in conformity with Glass2. They actually
state the above inequality at the glass transition in Eq. (10) of the above
paper but then take it to be also valid below the transition. Their conclusion
is based on the fact that $G\geq G_{\text{SCL}}$ for the Gibbs free energies
and the equality $H=H_{\text{SCL}}$ of the enthalpy at the glass transition;
here $S,G$ and $H$ refer to the glass and $S_{\text{SCL}},G_{\text{SCL}}$ and
$H_{\text{SCL}}$ refer to the supercooled liquid. The equality
$H=H_{\text{SCL}}$ is widely accepted as an experimental fact in the
field.\cite{Nemilov-Book,Debenedetti} The motivation for their proposal is
their understanding\cite{Kivelson,Gupta-JNCS,Gupta,Gupta1,Reiss} that the
entropy of a single sample of glass at absolute zero must be zero as the
microstate of the sample will not change no matter how long the glass is
"observed." Thus, according to this view, the entropy is not merely less than
that of Glass1, such as Glass3, it must be strictly less than or equal to
$S_{\text{SCL}}$. The alternative shown as Glass3%
\begin{equation}
S<S_{\text{Glass1}}>S_{\text{SCL}} \label{Alternative_GM}%
\end{equation}
is not consistent with this view.\cite{Gupta-JNCS}%

According to Eq. (\ref{GM_bound}), the entropy of a glass can only increase
during relaxation as the glass strives to equilibrate at any temperature below
the glass transition. What could be wrong with such a simple deduction, which
does not require any sophisticated mathematics? Gutzow and Schmelzer\cite[see
the discussion following their Eq. (39)]{Gutzow-Schmelzer} also come to the
same conclusion. On the other hand, the common view of the glass is shown by
Glass1 in Fig. \ref{Fig_entropyglass}, with the clear implication that the
entropy of the glass is higher than that of the supercooled liquid. There is
at present no consensus as is evident from the discussion reported in the
proceedings,\cite{Note1} and the general conclusion drawn by
Goldstein\cite{Goldstein} that the abrupt entropy loss comparable to
$S_{\text{R}}$ as proposed by Kivelson and Reiss\cite{Kivelson,Reiss} will
result in the construction of a perpetual motion machine of the second kind.
Goldstein merely follows the consequence of the existence of a possible
reversible connection between a glass and the supercooled liquid as proposed
by Kivelson and Reiss. In view of the simple observation in Eq.
(\ref{GM_bound}), one can argue that the general demonstration by Goldstein is
either inapplicable to a glass or that the basic premise of a possible
existence of a reversible connection between a glass and the supercooled
liquid using Kivelson-Reiss construction must be invalid. Indeed, Gupta and
Mauro\cite{Gupta,Gupta1} argue that the second law does not apply to the glass
transition proposed by them as the loss of entropy in their formulation does
not result in a latent heat, while Goldstein's general demonstration noted
above depends crucially on this heat and, therefore, does not invalidate the
proposal of a discontinuous loss by Gupta and Mauro although it does
invalidate the proposal by Kivelson and Reiss. One can also argue that the
special "electrode" construction proposed by Kivelson and Reiss, and used by
Goldstein, requires \emph{microscopic} information to confine the system into
a \emph{unique} basin, a configurational microstate, in that one needs to know
precisely where \emph{each} particle of the system is located within a certain
small volume cell associated with its possible
vibrations;\cite{Note-Measurement,Gujrati-book} see
Gujrati\cite{Gujrati-Symmetry} for elaboration on microstate measurements.
Oppenheim\cite{Oppenheim} has also raised somewhat of a similar objection. In
addition, it is not obvious that thermodynamics, which is a description of a
macrostate, can be applicable to a microstate such as the one obtained by the
above special construction.

Mauro et al\cite{Note1} also consider a glass formed by continuous cooling
such as Glass2 in which the entropy is lost continuously. It should be noted
that Goldstein\cite{Goldstein,Goldstein-1} does not explicitly discuss
continuous cooling; however, in a private discussion, Goldstein has pointed
out that his demonstration based on solubility also covers continuous cooling.
The amount of loss in this case is not going to be close to $S_{\text{R}}$.
Therefore, Goldstein's argument will not only invalidate Glass2 but also Glass3.

Despite all these attempts, the situation remains confusing, in part because,
to quote Goldstein\cite{Goldstein-1}\label{Goldstein-quote}: "As the residual
entropies found by calorimetric measurements either equal or at least do not
exceed the calculated value...., it is generally though not unanimously
accepted that the residual entropies are real." Thus, in these cases, as
already pointed out by Bestul and Chang,\cite{Betsul} one "...cannot
demonstrate..." whether the residual entropy "...differs from zero." Of
course, the attempts so far by those who believe in the residual
entropy,\cite{Jackel,Gutzow-Schmelzer,Nemilov,Johari,Sethna-Paper} have been
only to demonstrate that the residual entropy is \emph{approximately} equal to
the calculated value, to be denoted here by $S_{\text{expt}}(T_{0})$, within a
few percentages; however, as they do not deal directly with $S_{\text{SCL}}%
$,\ they have not ruled out the inequality in Eq. (\ref{GM_bound}). Even
though no experiments have revealed a situation in which $S_{\text{expt}%
}(T_{0})$ lies below $S_{\text{SCL}}(T_{0})$ over the temperature range where
the latter has been experimentally determined, the above calorimetric studies
make no claim that
\begin{equation}
S_{\text{expt}}(T_{0})>S_{\text{SCL}}(T_{0}) \label{Experiment_SCL_Bound}%
\end{equation}
for all temperatures below the glass transition, even though many workers
believe it to be intuitively true. In light of Eq. (\ref{GM_bound}), however,
t is possible that the above inequality turns out to be incorrect. The
contribution of Goldstein\cite{Goldstein} only proves that an abrupt or a
continuous loss of a part of the residual entropy ($S<S_{\text{Glass1}}$)
cannot be valid, but it neither rules out zero residual entropy nor the
inequality in Eq. (\ref{GM_bound}). It also says nothing about the above
inequality. Is it possible that the demonstration of $S\leq S_{\text{SCL}}$
itself is invalid? However, we cannot find any discussion of it in the current
literature. Thus, it is not surprising that the controversy still persists.
Goldstein\cite{Goldstein} has suggested that the entropy loss proposal "...may
be impossible to verify by any conceivable experiment,..." thus leaving the
possibility that some theoretical approach, such as the one to be taken here,
may resolve the issue. A theoretical approach will also allow us to
investigate the inequalties in Eqs. (\ref{GM_bound}-\ref{Experiment_SCL_Bound}%
) at absolute zero, where experiments cannot be performed.

\subsection{Clausius Limits\ \ \ \ }

Johari and Khouri\cite{Johari} have analyzed data from a large number of
glasses to draw the conclusion that a non-zero residual entropy is real in the
cases they have analyzed. The idea behind their discussion is simple: the
bounds, called the Clausius limits, \cite{Johari} on the (thermodynamic)
entropy reflect the contributions of irreversibility in the experimentally
determined estimate $S_{\text{expt}}(T_{0})$ of the entropy $S(T_{0})$. They
find that the two bounds are so tight in most cases (less than 2\%) that one
can neglect the corrections to the calorimetrically determined $S_{\text{expt}%
}(T_{0})$ due to irreversibility along cooling and heating paths, a conclusion
arrived at by several
others.\cite{Jackel,Gutzow-Schmelzer,Nemilov,Johari,GujratiResidualentropy} It
appears that this approach was first used by Bestul and Chang \cite{Betsul}
and later by Sethna and coworkers.\cite{Sethna-Paper}. Johari and Khouri
finally conclude that the experimental evidence of a non-zero residual entropy
(see Glass1 in Fig. \ref{Fig_entropyglass}) is beyond reproach for the systems
they have analyzed. The implication of their contribution and of several
others mentioned above\cite{Jackel,Gutzow-Schmelzer,Nemilov,Sethna-Paper} is
that there is no thermodynamic reason, such as the third law, for the residual
entropy to be zero, a conclusion well known in theoretical
physics.\cite{Landau} We should mention at this point that if there is ever
any \emph{conflict} between the second law \cite{note1} and any other law in
physics such as the zeroth or the third law for a macroscopic body, it is the
second law that is believed to hold in \emph{all} cases. This suggests that
Glass2 does not materialize for the systems considered by Johari and Khouri.
\cite{Johari}

\subsection{Importance of Bounds}

We wish to prove in this work that Glass2 cannot materialize for any system.
As far as Glass3 is concerned, we will establish that it may represent an
approximation for Glass1, but cannot represent a real glass. Therefore, we
will mostly consider Glass1 and Glass2 in the following. To fulfill this goal,
we have to go beyond the calorimetric
evidence\cite{Jackel,Gutzow-Schmelzer,Nemilov,Johari} for the residual
entropy. We treat the two glasses in Fig. \ref{Fig_entropyglass} basically as
two separate \emph{proposals} for the general behavior of glasses. Therefore,
they need to be demonstrated to be valid or invalid in \emph{all} cases,
without a single exception. To obtain such a general result, we must not
\emph{make} any assumptions or approximations; the latter would mean that the
conclusion could not be valid in all cases; see below. We must also not rely
upon heuristic or intuitive arguments as part of the proof. As the exact
values of the entropy require detailed information about the system, it in not
feasible to find the two entropies exactly for all systems. Thus, we will not
be interested in system-specific knowledge; rather, we want mathematically
sound conclusions about the entropy that will be valid for all glass formers
\emph{without exception}. Because the controversy is between two inequalities,
any approximation would turn the strict inequality such as $S<S_{\text{SCL}}$
into $S^{\text{approx}}\lessapprox S_{\text{SCL}}$. In that case, we may have
$S>S_{\text{SCL}}$in some cases depending on how far $S$ is from
$S^{\text{approx}}$. Unfortunately, various
discussions\cite{Jackel,Gutzow-Schmelzer,Nemilov,Johari} in favor of the
residual entropy and the specific critique by Goldstein\cite{Goldstein} using
solubility either use approximations (such as replacing the inequality in Eq.
(\ref{Residual_Entropy_determination}) below by an equality) or are
phenomenological with a limited domain of validity. Indeed, Gutzow and
Schmelzer\cite{Gutzow-Schmelzer} have come to the conclusion (see above Eq.
(40) there) that in any real process, the entropy either remains constant or
increases. The conclusion is in accordance with Eq. \ref{GM_bound}. Thus, the
current status of the field leaves open the possibility that the residual
entropy could vanish.

It is abundantly clear from the above discussion that there is a need to look
at the issue of residual entropy once again. In our opinion, the real issue is
the inequality in Eq. (\ref{GM_bound}), whose validity has not been
scrutinized in the literature. If the bound is jusitified, it automatically
proves that $S_{\text{R}}$ is not real. If, on the other hand, the bound is
not justified, it does not automatically prove that $S_{\text{R}}$ is real. We
then have the additional task to prove its reality. As is customary, we treat
the supercooled liquid as an equilibrium state, even though it not a true
equilibrium state; see above. We proceed by following the strict second law
inequality $d_{\text{i}}S>0$ in Eq. (\ref{Second_Law_Inequality}) and use it
along with the existence of internal equilibrium\cite{Guj-NE-I} (for Theorem
\ref{Theorem_Entropy_Variation}) to prove the following two theorems
applicable to \emph{all} non-equilibrium systems, regardless of how close or
far they are from their equilibrium state.

\begin{theorem}
\label{Theorem_Lower_Bound}The experimentally observed non-zero entropy at
absolute zero in a vitrification process is a \emph{strict lower bound of the
residual entropy} of any system:
\begin{equation}
S_{\text{R}}\equiv S(0)>S_{\text{expt}}(0)>S_{\text{SCL}}(0).
\label{ResidualEntropy_Bound}%
\end{equation}

\end{theorem}

\begin{theorem}
\label{Theorem_Entropy_Variation}Any drop of the glass entropy below that of
the supercooled liquid (such as Glass2 in Fig. \ref{Fig_entropyglass}) is a
violation of the second law. \cite{note1} Thus,
\begin{equation}
S>S_{\text{SCL}}, \label{GlassEntropy_Bound}%
\end{equation}
so that the entropy variation in time has a unique direction as shown by the
\emph{downward arrows} in Fig. \ref{Fig_entropyglass}.
\end{theorem}

Our conclusion in Eq. (\ref{GlassEntropy_Bound}) is in contradiction with that
in Eq. (\ref{GM_bound}). Therefore, we need to understand the reason for this
discrepancy. All experiments on or exact/approximate computations for
non-equilibrium systems \emph{must} obey the strict inequalities\ in Eqs.
(\ref{ResidualEntropy_Bound}-\ref{GlassEntropy_Bound}) without any exception.
This is the meaning behind the usage of "... rigorous ..." in the title of the
paper. The actual values of the entropy are not relevant for the aim of this
work, which is to settle the controversy between Glass1 and Glass2 under
vitrification and the way their entropies relate to that of the equilibrated
supercooled liquid. Because of the possibility that the systems may be far
away from equilibrium, where the irreversible contributions may not be
neglected, our results go beyond the previous calorimetric
evidence.\cite{Jackel,Gutzow-Schmelzer,Nemilov,Johari} The systems we are
interested in include glasses and imperfect crystals as special cases.
However, to be specific, we will only consider glasses below; the discussion
is valid for all non-equilibrium systems.

We hope that the proofs of the above two theorems settle the controversy
between the two forms of glasses and about the residual entropy $S_{\text{R}}$.

We first consider the behavior of the thermodynamic entropy that appears in
classical non-equilibrium thermodynamics. As is well known, this entropy is
governed by the second law.\cite{note1} In the next section, we prove Theorem
\ref{Theorem_Lower_Bound}. In the following section, we prove Theorem
\ref{Theorem_Entropy_Variation}. Both sections deal with the thermodynamic
entropy. In Sect. \ref{Sect_Stat_Entropy}, we explain how the thermodynamic
residual entropy can be understood in terms probability of a microstate, and
how even a single sample can give a highly reliable result for the residual
entropy.\textbf{ }The last section contains our conclusions.

\section{Forward Entropy Bound during Vitrification:\ Theorem
\ref{Theorem_Lower_Bound}}

The process we consider is carried out at some cooling rate as follows. The
temperature of the medium is isobarically changed by some small but fixed
$\Delta T_{0}$ from the current value to the new value, and we wait for (not
necessarily fixed) time $\tau_{\text{obs}}$ at the new temperature to make an
instantaneous measurement on the system before changing the temperature again.
At some temperature $T_{0\text{g}}$, see Fig. \ref{Fig_entropyglass}, the
relaxation time $\tau_{\text{relax}}$, which continuously increases as the
temperature is lowered, becomes equal to $\tau_{\text{obs}}$. Just below
$T_{0\text{g}}$, the structures are not yet frozen; they "freeze" at a lower
temperature $T_{0\text{G}}$ (not too far from $T_{0\text{g}})$ to form an
amorphous solid with a viscosity close to $10^{13}$ poise. This solid is
identified as a \emph{glass}. The location of both temperatures depends on the
rate of cooling, i.e. on $\tau_{\text{obs}}$. Over the glass transition region
between $T_{0\text{G}}$ and $T_{0\text{g}}$\ shown in Fig.
\ref{Fig_entropyglass}, the \emph{non-equilibrium} liquid gradually turns from
an equilibrium supercooled liquid at or above $T_{0\text{g}}$ into a glass at
or below $T_{0\text{G}}$, a picture already known since
Tammann.\cite{Nemilov-Book} Over this region, some dynamical properties such
as the viscosity vary continuously but very rapidly. However, thermodynamic
quantities such as the volume or the enthalpy change continuously but slowly.
As the observation time $\tau_{\text{obs}}$ is increased, the equilibrated
supercooled liquid continues to lower temperatures before the appearance of
$T_{0\text{g}}$. In the \emph{hypothetical limit} $\tau_{\text{obs}%
}\rightarrow\infty$, it is believed that the equilibrated supercooled liquid
will continue to lower temperatures without any interruption, and is shown
schematically by the solid blue curve in Fig. \ref{Fig_entropyglass}. We
overlook the possibility of the supercooled liquid ending in a
spinodal.\cite{Gujrati-spinodal} It is commonly believed that this entropy
will vanish at absolute zero ($S_{\text{SCL}}(0)\equiv0$), as shown in the
figure. As we are going to be interested in $S_{\text{SCL}}(T_{0})$ over
$(0,T_{0\text{g}})$, we must also acknowledge the possibility of an ideal
glass transition in the system. If one believes in an ideal glass transition,
then there would be a singularity in $S_{\text{SCL}}(T_{0})$ at some positive
temperature $T_{\text{K}}<T_{0\text{G}},$ below which the system will turn
into an ideal glass whose entropy will also vanish at absolute
zero.\cite{Nemilov-Book} The possibility of an ideal glass transition, which
has been discussed in a recent review elsewhere,\cite{Gujrati-book} will not
be discussed further in this work. All that will be relevant in our discussion
here is the fact that the entropy vanishes in both situations ($S_{\text{SCL}%
}(0)\equiv0$). However, it should be emphasized that the actual value of
$S_{\text{SCL}}(0)$ has no relevance for the theorems.

It is a common practice to think of the glass transition to occur at a point
that lies between $T_{0\text{g}}$ and $T_{0\text{G}}$.\ Gupta and
Mauro\cite{Gupta-JNCS} consider the glass transition to occur at
$T_{0\text{G}}$\ to obtain the bound in Eq. (\ref{GM_bound}). We will not make
this assumption in this work except when we discuss their inequality later. We
have drawn the two entropy curves (Glass1 or Glass2) in Fig.
\ref{Fig_entropyglass} that emerge out of $S_{\text{SCL}}(T_{0})$ for a given
$\tau_{\text{obs}}$ in such a way that Glass1 has its entropy above (so that
$S_{\text{R}}\geq0$)\ and Glass2 below (so that $S_{\text{R}}\equiv0$) that of
the supercooled liquid. The entropy of Glass1 (Glass2) approaches that of the
equilibrated supercooled liquid entropy from above (below) during isothermal
(fixed temperature of the medium) relaxation; see the two downward vertical
arrows for Glass1. It is the approach to equilibrium that distinguishes the
two glasses, Glass1 and Glass2. Although Johari and Khouri do not mention in
their conclusion, their analysis of tight bounds also shows that the entropy
\emph{does not} drop by an amount close to $S_{\text{R}}$ within the glass
transition region for the systems studied by them. However, because of the
involved approximation, it sheds no light on whether the glass entropy lies
above or below the corresponding $S_{\text{SCL}}(T_{0})$. Thus, their work and
many others leave open the possibility that Glass2 may materialize if the
irreversibility is too large. This again shows why obtaining a bound is so
important, even if we do not determine the actual entropy values.

The concept of internal equilibrium\cite{Guj-NE-I,Guj-NE-II} for a
non-equilibrium system means that its instantaneous entropy is a state
function of its instantaneous state variables like energy, volume etc. and any
internal variables\cite{Donder,deGroot,Prigogine,Guj-NE-I,Guj-NE-II} used to
specify its state. Their usage is also a common practice
\cite{Gutzow-book,Nemilov-Book} now-a-days for glasses. Employing internal
equilibrium gives rise to an instantaneous Gibbs fundamental relation, see Eq.
(\ref{Gibbs_Fundamental_Equation}) below, which determines its instantaneous
temperature, pressure, etc.

We now prove Theorem \ref{Theorem_Lower_Bound}. Consider an isobaric process
(we will not explicitly exhibit the pressure in this section) from some state
A at temperature $T_{0}$ in the supercooled liquid state which is still higher
than $T_{0\text{g}}$ to the state A$_{0}$ at absolute zero. The state A$_{0}$
depends on the path A$\rightarrow$A$_{0}$ along with $T_{0}=0$, which is
implicit in the following. We have along A$\rightarrow$A$_{0}$%
\begin{equation}
S(0)=S(T_{0})+%
{\textstyle\int\limits_{\text{A}}^{\text{A}_{0}}}
d_{\text{e}}S+%
{\textstyle\int\limits_{\text{A}}^{\text{A}_{0}}}
d_{\text{i}}S, \label{General_Entropy_Calculation}%
\end{equation}
where we have assumed that there is no latent heat in the vitrification
process, and where \cite{Donder,deGroot,Prigogine,Guj-NE-I,Guj-NE-II}
$dS=d_{\text{e}}S+d_{\text{i}}S$, each of which for a non-equilibrium system
is path dependent. The component
\begin{equation}
d_{\text{e}}S(t)=-dQ(t)/T_{0}\equiv C_{P}dT_{0}/T_{0}
\label{Heat_Capacity_Relation}%
\end{equation}
represents the reversible entropy exchange with the medium in terms of the
heat $dQ(t)$ given out by the glass at time $t$ to the medium whose
temperature at that instant is $T_{0}$. The component $d_{\text{i}}S>0$
represents the irreversible entropy generation in the irreversible process. In
general, the \emph{irreversible} term%
\begin{equation}
d_{\text{i}}S\geq0 \label{Second_Law_Inequality}%
\end{equation}
contains, in addition to the contribution from the irreversible heat transfer
with the medium, contributions from all sorts of viscous dissipation going on
\emph{within} the system and normally require the use of internal
variables.\cite{Donder,deGroot,Prigogine,Guj-NE-I,Guj-NE-II} Thus, Eq.
(\ref{General_Entropy_Calculation}) contains \emph{all} possible sources of
entropy variations. \cite{Donder,deGroot,Prigogine,Guj-NE-I,Guj-NE-II} This is
easily proven by considering the system and the medium as an \emph{isolated
system}\cite{Guj-NE-I,Guj-NE-II} and expressing the entropies as functions of
the instantaneous values of the \emph{observables} and \emph{ internal
variables}. A discontinuous change in the entropy is ruled out from the
continuity of the Gibbs free energy $G$ and the enthalpy $H$ in vitrification
proved elsewhere.\cite{Guj-NE-I} Thus, we only consider a continuous change in
the entropy as shown by the two glass curves in Fig. \ref{Fig_entropyglass}.

The equality in Eq. (\ref{Second_Law_Inequality}) holds for a reversible
process, which we will no longer consider unless stated otherwise. The strict
inequality $d_{\text{i}}S>0$ occurs only for irreversible process such as in a
glass. Since the second integral in Eq. (\ref{General_Entropy_Calculation}) is
always \emph{positive}, and since the residual entropy $S_{\text{R}}$ is, by
definition, the entropy $S(0)$ at absolute zero, we obtain the important
result%
\begin{equation}
S_{\text{R}}\equiv S(0)>S_{\text{expt}}(0)\equiv S(T_{0})+%
{\textstyle\int\limits_{T_{0}}^{0}}
C_{P}dT_{0}/T_{0}.\label{Residual_Entropy_determination}%
\end{equation}
This confirms the expectation noted above that the irreversibility during
vitrification does not allow for the determination of the entropy exactly,
because determining the second integral in Eq.
(\ref{General_Entropy_Calculation}) is not trivial, especially if internal
variables are not considered. \cite{Guj-NE-II,Nemilov-Book} The forward
inequality is due to the irreversible entropy generation from all possible
sources\cite{Donder,deGroot,Prigogine,Guj-NE-I,Guj-NE-II} that seems to not
have been recognized by the
proponents\cite{Reiss,Gupta-JNCS,Kivelson,Gupta,Gupta1} of vanishing
$S_{\text{R}}$. This strict forward inequality clearly establishes that the
residual entropy at absolute zero must be strictly larger than $S_{\text{expt}%
}(0)$ in any non-equilibrium process. This proves the first inequality in Eq.
(\ref{ResidualEntropy_Bound}).

We now prove the second inequality in Eq. (\ref{ResidualEntropy_Bound}). We
consider processes that occur when $\tau_{\text{obs}}<\tau_{\text{relax}%
}(T_{0})$. Let $\overset{\cdot}{Q}(t)\equiv dQ(t)/dt$ be the rate of net heat
loss by the system. Then, for each temperature interval $dT_{0}<0$ below
$T_{0\text{g}}$, we have
\[
\left\vert dQ\right\vert \equiv C_{P}\left\vert dT_{0}\right\vert =%
{\textstyle\int\limits_{0}^{\tau_{\text{obs}}}}
\left\vert \overset{\cdot}{Q}\right\vert dt<\left\vert dQ\right\vert
_{\text{eq}}(T_{0})\equiv%
{\textstyle\int\limits_{0}^{\tau_{\text{relax}}(T_{0})}}
\left\vert \overset{\cdot}{Q}\right\vert dt,\ \ \ T_{0}<T_{0\text{g}}%
\]
where $\left\vert dQ\right\vert _{\text{eq}}(T_{0})>0$ denotes the net heat
loss by the system to come to equilibrium, i.e. become supercooled liquid
during cooling at $T_{0}$. For $T_{0}\geq T_{0\text{g}}$, we have $dQ\equiv
dQ_{\text{eq}}(T_{0})$. Thus,%
\[%
{\textstyle\int\limits_{T_{0}}^{0}}
C_{P}dT_{0}/T_{0}>%
{\textstyle\int\limits_{T_{0}}^{0}}
C_{P\text{,eq}}dT_{0}/T_{0},
\]
where $\left\vert dQ\right\vert _{\text{eq}}\equiv C_{P\text{,eq}}\left\vert
dT_{0}\right\vert $. We thus conclude that
\begin{equation}
S_{\text{expt}}(0)>S_{\text{SCL}}(0); \label{Entropy_bound_at_0}%
\end{equation}
the strict inequality is the result of the fact that glass is a
non-equilibrium state. Otherwise, we will have $S_{\text{expt}}(0)\geq
S_{\text{SCL}}(0)$ for any arbitary state.

This proves Theorem \ref{Theorem_Lower_Bound}.

The difference $S_{\text{R}}-$ $S_{\text{expt}}(0)$ would be larger, more
irreversible the process is. The quantity $S_{\text{expt}}(0)$ can be
determined calorimetrically by performing a cooling experiment. We take
$T_{0}$ to be the melting temperature $T_{0\text{M}}$, and uniquely determine
the entropy of the supercooled liquid at $T_{0\text{M}}$ by adding the entropy
of melting to the crystal entropy $S_{\text{CR}}(T_{0\text{M}})$ at
$T_{0\text{M}}$. The latter is obtained in a unique manner by integration
along a reversible path from $T_{0}=0$ to $T_{0}=T_{0\text{M}}$:
\[
S_{\text{CR}}(T_{0\text{M}})=S_{\text{CR}}(0)+%
{\textstyle\int\limits_{0}^{T_{0\text{M}}}}
C_{P\text{,CR}}dT_{0}/T_{0},
\]
here, $S_{\text{CR}}(0)$ is the entropy of the crystal at absolute zero, which
is traditionally taken to be zero in accordance with the third law, and
$C_{P\text{,CR}}(T_{0})$ is the isobaric heat capacity of the crystal. This
then uniquely determines the entropy of the liquid to be used in the right
hand side in Eq. (\ref{Residual_Entropy_determination}). We will assume that
$S_{\text{CR}}(0)=0$. Thus, the experimental determination of $S_{\text{expt}%
}(0)$ is required to give the \emph{lower bound} to the residual entropy in
Eq. (\ref{ResidualEntropy_Bound}). Experiment evidence for a non-zero value of
$S_{\text{expt}}(0)$ is abundant as discussed by several
authors;\cite{Giauque,Giauque-Gibson,Jackel,Gutzow-Schmelzer,Nemilov,Goldstein}
various textbooks\cite{Gutzow-book,Nemilov-Book} also discuss this issue.
Goldstein\cite{Goldstein} gives a value of $S_{\text{R}}\simeq15.1$ J/K mol
for \textit{o-}terphenyl from the value of its entropy at $T_{0}=2$ K.
However, we have given a mathematical justification of $S_{\text{expt}}%
(0)>0$\ in Eq. (\ref{Entropy_bound_at_0}). The strict inequality proves
immediately that the residual entropy \emph{cannot} vanish for glasses, which
justifies the curve Glass1 in Fig. \ref{Fig_entropyglass}.

The inequality in Eq. (\ref{Residual_Entropy_determination}) takes into
account any amount of irreversibility during vitrification; it is no longer
limited to only small contributions of the order of $2\%$ considered by Johari
and Khouri and by several
others,\cite{Gutzow-book,Nemilov-Book,GujratiResidualentropy,Goldstein} which
makes our derivation very general. The relevance of the residual entropy has
been discussed by several authors in the literature.
\cite{Pauling,Tolman,Sethna,Chow,Goldstein,GujratiResidualentropy,Gujrati-Symmetry,Gutzow-Schmelzer,Nemilov}%

By considering the state A$_{0}$ above to be a state A$_{0}$\ of the glass in
a medium at some arbitrary temperature $T_{0}^{\prime}$ below $T_{0\text{g}}$,
we can get a generalization of Eq. (\ref{Residual_Entropy_determination}):%
\begin{equation}
S(T_{0}^{\prime})>S_{\text{expt}}(T_{0}^{\prime})\equiv S(T_{0})+%
{\textstyle\int\limits_{T_{0}}^{T_{0}^{\prime}}}
C_{P}dT_{0}/T_{0}. \label{Entropy_determination}%
\end{equation}
We again wish to remind the reader that all quantities depend on the path
A$\rightarrow$A$_{0}$, which we have not exhibited. By replacing $T_{0}$\ by
the melting temperature $T_{0\text{M}}$ and $T_{0}^{\prime}$\ by $T_{0}$, and
adding the entropy $\widetilde{S}(T_{0\text{M}})$ of the medium on both sides
in the above inequality, and rearranging terms, we obtain (with $S_{\text{L}%
}(T_{0\text{M}})=S_{\text{SCL}}(T_{0\text{M}})$ for the liquid)%
\begin{equation}
S_{\text{L}}(T_{0\text{M}})+\widetilde{S}(T_{0\text{M}})\leq S(T_{0}%
)+\widetilde{S}(T_{0\text{M}})-%
{\textstyle\int\limits_{T_{0\text{M}}}^{T_{0}}}
C_{P}dT_{0}/T_{0}, \label{Setna_Inequality}%
\end{equation}
where we have also included the equality for a reversible process. This
provides us with an independent derivation of the inequality given by Setna
and coworkers.\cite{Sethna-Paper}

It is also clear from the derivation of Eq. (\ref{Entropy_bound_at_0}) that
the inequality can be generalized to any temperature $T_{0}<T_{0\text{g}}$
with the result%
\begin{equation}
S_{\text{expt}}(T_{0})>S_{\text{SCL}}(T_{0}), \label{Entropy_bound}%
\end{equation}
with $S_{\text{expt}}(T_{0})\rightarrow S_{\text{SCL}}(T_{0})$ as
$T_{0}\rightarrow T_{0\text{g}}$ from below. Thus, $S_{\text{expt}}(T_{0})$
appears in form similar to that of Glass3 in Fig. \ref{Fig_entropyglass},
except that the latter represents a possible glass entropy while the former
represents the calorimetric approximation for Glass1.

While we have only demonstrated the forward inequality, the excess
$S_{\text{R}}-S_{\text{expt}}(0)$ can be computed in non-equilibrium
thermodynamics,\cite{Donder,deGroot,Prigogine,Guj-NE-I,Guj-NE-II} which
provides a clear prescription for calculating the irreversible entropy
generation. We do not do this here as we are only interested in general
results, while the calculation of irreversible entropy generation will, of
course, be system-dependent and will require detailed information. Gutzow and
Scmelzer\cite{Gutzow-Schmelzer} provide such a procedure with a single
internal variable but under the assumption of equal temperature and pressure
for the glass and the medium. However, while they comment that $d_{\text{i}%
}S\geq0$ whose evaluation requires system-dependent properties, their main
interest is to only show that it is negligible compared to $d_{\text{e}}S$.

We have proved Theorem \ref{Theorem_Lower_Bound} by considering only the
system without paying any attention to the medium but assuming the second law
as is evident from Eq. (\ref{Second_Law_Inequality}). We have done this
because the proponents of vanishing $S_{\text{R}}$ normally consider the
glassy state without ever bringing in the medium in the discussion. This does
not mean that the conclusion would be any different had we brought the medium
into our discussion. This is seen from the derivation of the ineqaulity in Eq.
(\ref{Setna_Inequality}) from Eq. (\ref{Entropy_determination}). We will find
it convenient to consider the medium in the next section to overcome the
objection\cite{Gupta,Gupta1} that the glass does not obey the second law.

\section{Entropy and Enthalpy during Relaxation}

We now turn to the inequality in Eq. (\ref{GM_bound}) to see if it would ever
be satisfied. To avoid directly discussing the relationship of the latent heat
with the entropy loss, the possibility of entropy loss, or whether the second
law applies to a glass, we change our mode of presentation and consider the
system not by itself, but as a part of an isolated system in which the system
is surrounded by an \emph{extremely large }medium whose temperature $T_{0}$,
pressure $P_{0}$, etc. are not affected by what happens within the system. To
prove Theorem \ref{Theorem_Entropy_Variation}, we consider the system to be
not in equilibrium with the medium. All processes that go on within the medium
occur at constant temperature, pressure, etc. Thus, there will not be any
irreversible process going on within the medium. All irreversible processes
will go on within the system. This simplification occurs because of the
extremely large size of the medium and will be central in our discussion here.

While some may doubt that the second law is not applicable to the glass at the
glass transition, there cannot be any doubt that the second law applies to the
isolated system. It is found \cite{Guj-NE-I,Guj-NE-II,Langer} that the
instantaneous temperature $T(t)$, pressure $P(t)$, etc. of the system are
different from the corresponding quantities of the medium when the former is
not in equilibrium with the medium. All that is required is for the system to
be in \emph{internal equilibrium},\cite{Guj-NE-I,Guj-NE-II} which is defined
as a state in which the entropy has no explicit time-dependence; its time
variation is due to time-dependent observables and internal variables. The
existence of instantaneous $T(t)$, $P(t)$, etc. is a consequence of internal
equilibrium, and is a general property of any system out of equilibrium even
at high temperatures and is not restricted to glasses only. Therefore, $T(t)$,
$P(t)$, etc. should not be confused with fictive temperature and pressure,
etc. that are meaningful for glasses. This issue has been discussed
earlier.\cite{Guj-NE-I} The Gibbs fundamental relation for the system when it
is in internal equilibrium is given by%
\begin{equation}
dE(t)=T(t)dS(t)-P(t)dV(t)-A(t)d\xi(t), \label{Gibbs_Fundamental_Equation}%
\end{equation}
where we have allowed for a single internal or structure variable $\xi(t)$ for
the sake of simplicity; for glasses, see Nemilov\cite{Nemilov-Book} and Gutzow
and Schmelzer\cite{Gutzow-book} for the usage of internal variables. We have
explicitly shown the time-dependence in the above equation to highlight the
presence of relaxation in the system. The affinity $A(t)$ is conjugate to the
internal variable and vanishes when the system comes to equilibrium with the
medium ($A_{0}\equiv0$). In that case, we also have $T(t)\rightarrow T_{0}$
and $P(t)\rightarrow P_{0}$; see Eq.
(\ref{Gibbs_Fundamental_Equation_Expanded}).

For a system out of equilibrium, the instantaneous entropy $S(t)$ and volume
$V(t)$ seem to play the role\cite{Guj-NE-I} of "internal variables," whose
"affinities" are given by $T(t)-T_{0}$ and $P(t)-P_{0}$, respectively. This
fact is not common in the glass literature to the best of our knowledge. The
temperature and pressure of the system are usually taken to be those of the
medium, which is an approximation. For example, Schmelzer and
Gutzow\cite{Gutzow-1} identify $d_{\text{e}}S=C_{P}dT/T$, see their Eq. (1),
whereas it should be properly identified as in Eq.
(\ref{Heat_Capacity_Relation}) with $T$ replaced by $T_{0}$. They also
identify the pressure in their Gibbs fundamental relation [see their Eq. (2)]
as the external pressure.

We now turn to prove Theorem \ref{Theorem_Entropy_Variation}. Let us rewrite
the Gibbs fundamental relation as
\begin{equation}
dE(t)=T_{0}dS(t)-P_{0}dV(t)+(T(t)-T_{0})dS(t)-(P(t)-P_{0})dV(t)-A(t)d\xi(t),
\label{Gibbs_Fundamental_Equation_Expanded}%
\end{equation}
in which each of the last three terms can be associated with an irreversible
entropy generation.\cite{Guj-NE-I} For this, it is easier to take all but the
first term on the right side to the other side of the equation. We thus
note\cite{Guj-NE-I} that
\begin{equation}
(T_{0}-T(t))dS(t)\geq0,(P(t)-P_{0})dV(t)\geq0,A(t)d\xi(t)\geq0,
\label{Second_Law_Inequalities}%
\end{equation}
in accordance with the second law. The equalities above and below occur only
for reversible processes. As we are only interested in irreversible processes
in non-equilibrium systems, the inequalities above and below become strict
inequalities, which cannot be violated in any real process. Thus, as before,
we will exploit these \emph{strict} inequalities to derive a bound on the rate
of entropy variation.

We extend the derivation given earlier\cite{Guj-NE-I} to include the internal
variable to obtain as the statement of the second law:\cite{note1}%
\begin{equation}
\frac{dS_{0}(t)}{dt}=\left(  \frac{1}{T(t)}-\frac{1}{T_{0}}\right)
\frac{dE(t)}{dt}+\left(  \frac{P(t)}{T(t)}-\frac{P_{0}}{T_{0}}\right)
\frac{dV(t)}{dt}+\frac{A(t)}{T(t)}\frac{d\xi(t)}{dt}\geq0;
\label{Total_Entropy_Rate}%
\end{equation}
each term in the first equation must be non-negative. In a vitrification
process, in which the energy decreases with time, we must, therefore, have%
\[
T(t)\geq T_{0}%
\]
during any relaxation (at a fixed temperature and pressure of the medium) so
that $T(t)$ approaches $T_{0}$ from above [$T(t)$ $\rightarrow$ $T_{0}^{+}$]
as the relaxation ceases and the equilibrium is achieved. It now follows from
Eq. (\ref{Second_Law_Inequalities}) that during vitrification%
\begin{equation}
dS(t)/dt\leq0; \label{Entropy_variation}%
\end{equation}
the equality occurring only when equilibrium with the medium has been
achieved. The above inequality gives the bound on the rate that we are
interested in. At the end of the relaxation%
\[
S(T_{0},P_{0},t)\overset{t\rightarrow\infty}{\rightarrow}S_{\text{SCL}}%
^{+}(T_{0},P_{0});
\]
the plus sign is to indicate that the glass entropy reaches $S_{\text{SCL}%
}(T_{0},P_{0})$ from above.

We have shown $T_{0},P_{0}$ in $S(T_{0},P_{0},t)\equiv S(T(t),P(t),A(t))$ to
emphasize that the result is general during any relaxation. In the derivation,
which only considers the behavior of $S_{0}(t)$ of the isolated system,\ no
assumption about the nature of irreversibility such as any loss of ergodicity
in the system, inapplicability of the second law to the system, possibility of
any chaotic behavior, chemical reaction, etc. is made. The only assumption
that has been made is that it is possible to define the instantaneous
temperature and pressure for the system. We have also made no assumption that
$S(t)$ lies above (Glass1,\ Glass3) or below (Glass2) the entropy
$S_{\text{SCL}}(T_{0})$ of the equilibrated supercooled liquid; see Fig.
\ref{Fig_entropyglass}. Being a general result, it should be valid for any
real glass. This now gives a way to decide which of the glasses in Fig.
\ref{Fig_entropyglass} is in accordance with the above conclusion. Above
$T_{0\text{g}}$, the system is always in equilibrium with the medium so its
temperature is the same as $T_{0}$. Below $T_{0\text{g}}$, when the system is
not in equilibrium with the medium, then $T(t)>T_{0}$ as long as there is no
equilibrium. The entropy of Glass1 and Glass3 approach $S_{\text{SCL}}(T_{0})$
of the equilibrated supercooled liquid from above during any isothermal
relaxation, which is consistent with Eq. (\ref{Entropy_variation}). As the
entropy is a unique function of the path, the two glasses must correspond to
different histories. Therefore, from now on, we will not consider Glass3 anymore.

For Glass2 in Fig. \ref{Fig_entropyglass}, the entropy actually drops below
that of the supercooled liquid by some
amount;\cite{Reiss,Gupta-JNCS,Jackel,Palmer,Hemmen,Thirumalai,Kivelson,Gupta}
the amount does not even have to be comparable to $S_{\text{R}}$. In this
situation, the entropy must approach $S_{\text{SCL}}(T_{0})$ of the
supercooled liquid from below during relaxation; see the upward arrow in Fig.
\ref{Fig_entropyglass}. This will result in the increase of the entropy during
relaxation, which violates Eq. (\ref{Entropy_variation}). Thus, Glass2 cannot
be rationalized. For the same reason, the conclusion of Gutzow and
Schmelzer\cite{Gutzow-Schmelzer} of increase in entropy cannot be rationalized.

We are now ready to investigate what could be technically wrong with the
entropy loss proposal. This will also settle whether the glass must obey the
second law or not at the glass transition. From the behavior of $dS_{0}(t)/dt$
in Eq. (\ref{Total_Entropy_Rate}), we can immediately
identify\cite{Guj-NE-I,Guj-NE-II} the rates for the entropy of the system and
the medium, respectively,%
\begin{align*}
\frac{dS(t)}{dt}  &  =\frac{1}{T(t)}\frac{dE(t)}{dt}+\frac{P(t)}{T(t)}%
\frac{dV(t)}{dt}+\frac{A(t)}{T(t)}\frac{d\xi(t)}{dt}.\\
\frac{d\widetilde{S}(t)}{dt}  &  =-\frac{1}{T_{0}}\frac{dE(t)}{dt}-\frac
{P_{0}}{T_{0}}\frac{dV(t)}{dt}.
\end{align*}
While the entropy change of the medium has no irreversible contribution as
noted earlier, the irreversible entropy change $d_{\text{i}}S(t)$ of the
system is given by the three terms in Eq. (\ref{Total_Entropy_Rate}), each of
which must be non-negative. Writing $dS(t)=d_{\text{e}}S(t)+d_{\text{i}}S(t)$,
we find that
\begin{align*}
\frac{d_{\text{e}}S(t)}{dt}  &  =-\frac{d\widetilde{S}(t)}{dt}=\frac{1}{T_{0}%
}\frac{dE(t)}{dt}+\frac{P_{0}}{T_{0}}\frac{dV(t)}{dt}\\
\frac{d_{\text{i}}S(t)}{dt}  &  =\frac{dS_{0}(t)}{dt}\geq0\text{.}%
\end{align*}
In general, $d_{\text{e}}S(t)/dt$\ can have either sign. In a cooling process,
we have $dE(t)/dt<0$. Moreover, we normally have $dV(t)/dt<0$. Thus,
$d_{\text{e}}S(t)/dt<0$. However, its sign is not relevant for our discussion.
On the other hand, $d_{\text{i}}S(t)/dt\geq0$, which follows from the second
law applied to the isolated system, for which there is no dispute. With these
consequences of the second law, we can now evaluate the merit of the entropy
loss proposal.\cite{Reiss,Gupta-JNCS,Kivelson,Gupta,Gupta1} Let us consider
vitrification. The change in the entropy $d_{\text{e}}S(t)<0$ is due to
exchanges with the medium. This is part of the entropy change that will occur
even in a reversible process. The contribution from the entropy
loss,\cite{Reiss,Gupta-JNCS,Kivelson,Gupta,Gupta1} which we denote by
$dS_{\text{loss}}(t)$, is due to the vitrification process. Vitrification in
the entropy loss view represents changes occurring \emph{within} the system.
As the contribution from every internal process must be included in
$d_{\text{i}}S(t)$, and as each such contribution must be \emph{non-negative},
there is no way to justify the negative contribution $dS_{\text{loss}}(t)$ due
to vitrification\cite{Reiss,Gupta-JNCS,Kivelson,Gupta,Gupta1} without
violating the second law for the isolated system. This proves Theorem
\ref{Theorem_Entropy_Variation}.

This is the conclusion obtained by Goldstein;\cite{Goldstein} we have just
provided a direct proof of his conclusion. There is also no merit to the
suggestion of Gupta and Mauro\cite{Gupta-JNCS,Gupta,Gupta1} that the glass
does not obey the second law.

It is easy to see that the discussion above can be easily extended to include
other observables and internal variables without affecting Theorem
\ref{Theorem_Entropy_Variation}. The above discussion was also not restricted
to a constant pressure of the medium. Indeed, the discussion above has been
very general. The only restriction was the extremely large size of the medium,
which is easily satisfied in experiments. For the general case, the Gibbs free
energy is given by $G(t)=H(t)-T_{0}S(t)$. At fixed $T_{0}$, we
have\cite{Guj-NE-I}
\begin{equation}
\frac{dG(t)}{dt}=\frac{dH(t)}{dt}-T_{0}\frac{dS(t)}{dt}\leq0,
\label{Total_Entropy_Rate_1}%
\end{equation}
from which it follows that%
\[
\left\vert \frac{dH(t)}{dt}\right\vert \geq T_{0}\left\vert \frac{dS(t)}%
{dt}\right\vert .
\]
We now turn to considering isobaric vitrification during which the system is
not in equilibrium with the medium. We will assume that the system is always
very close to mechanical equilibrium so that its pressure is equal to $P_{0}$;
however, there is normally no thermal equilibrium so that the instantaneous
temperature $T(t)$ of the system is different from $T_{0}$. We will now show
that the above general conclusion remains unaltered. From Eq.
(\ref{Gibbs_Fundamental_Equation}), we have
\begin{equation}
\left\vert \frac{dH(t)}{dt}\right\vert \geq T(t)\left\vert \frac{dS(t)}%
{dt}\right\vert , \label{Entropy_variation1}%
\end{equation}
where we have used $A(t)d\xi(t)/dt\geq0$. The last bound is tighter than the
previous bound and reduces to the equality obtained earlier\cite{Guj-NE-I} in
the absence of any internal variable $\xi$. In any case, the enthalpy of the
glass is not constant in time at $T_{0\text{G}}$ even for an isobaric
vitrification if we accept that the entropy undergoes relaxation. We thus
conclude that there is no justification in assuming $\Delta H(T_{0\text{G}%
})=0$.

We now turn to the inequality in Eq. (\ref{GM_bound}) to inquire what may be
wrong with its trivial justification offered by its authors. We consider
$T_{0}\leq$ $T_{0\text{G}}$. The enthalpy $H(T_{0})$ of the glass relaxes
towards its equilibrium value $H_{\text{SCL}}(T_{0})$ of the supercooled
liquid from above during isobaric vitrification, contrary to the assumption by
Gupta and Mauro.\cite{Note1} Acknowledging this immediately leads to%
\[
T_{0\text{G}}(S(T_{0\text{G}})-S_{\text{SCL}}(T_{0\text{G}}))\leq
H(T_{0\text{G}})-H_{\text{SCL}}(T_{0\text{G}}).
\]
As the right side is strictly positive, there is no justification in
concluding the inequality in Eq. (\ref{GM_bound}). It is most certainly
possible to satisfy the above inequality and also have%
\[
S(T_{0\text{G}})\geq S_{\text{SCL}}(T_{0\text{G}}).\text{\ \ }%
\]
This is consistent with strict inequality in Eq. (\ref{Entropy_variation}).
The equality $\Delta H(T_{0})=0$ is only valid at $T_{0\text{g}}$, where the
difference $\Delta S(T_{0})=0$, so that $S(T_{0\text{g}})=S_{\text{SCL}%
}(T_{0\text{g}})$.

The isothermal relaxation that occurs during glassy vitrification originates
from the tendency of the glass to come to thermal equilibrium during which its
temperature $T(t)$ approaches $T_{0}$ in time. The relaxation process results
in the lowering of the corresponding Gibbs free energy \cite{Guj-NE-I} in
time, as expected; this results in not only lowering the enthalpy during
vitrification, as observed experimentally, but also of the entropy $S(t)$
during relaxation, as shown for Glass1 in Fig. \ref{Fig_entropyglass}.

\section{Statistical Interpretation of Thermodynamic $S_{\text{R}}%
$\label{Sect_Stat_Entropy}}

Now that we have established the reality of the residual entropy $S_{\text{R}%
}$ by considering the thermodynamic entropy in classical thermodynamics, we
wish to discuss its possible statistical interpretation. We recall that the
thermodynamic entropy cannot be given a unique value; all that we can discuss
in thermodynamics is the change in it. It does not even have to be
non-negative, as is evident from the entropy of an ideal gas at very low
temperature. However, as we have defined $S(T_{0},t)$ with respect to
$S_{\text{CR}}(0)$ in this work, and we have taken $S_{\text{CR}}(0)=0$, the
thermodynamic entropy has a \emph{unique} value. Therefore, any attempt to
provide a statistical interpretation must result in an agreement with the
numerical value of the above thermodynamic entropy, a point also made by
Sethna and coworkers.\cite{Sethna-Paper} The issue of the statistical
interpretation has been discussed elsewhere by
us.\cite{GujratiResidualentropy,Gujrati-Symmetry} As a glass is a frozen
structure (over an extremely long period of time), we index each frozen
structure, which represents a microstate, at absolute zero by $i=1,2,\cdots
,W_{\text{G}}$. Each microstate is characterized by the set of observables and
internal variables. All glasses formed under\emph{ identical macroscopic
conditions} will be in one of these microstates at $T_{0}=0$; let $p_{i}>0$
denote the probability that a glass will be in the microstate $i$. The
instantaneous statistical entropy is an average
quantity\cite{Gujrati-Symmetry,Landau} (we set $k_{\text{B}}=1$%
):$S(t)=-\left\langle \ln p\right\rangle \equiv\sum_{i}p_{i}(-\ln p_{i}).$ If
the glass formation occurs under an unbiased condition, all microstates will
be equally probable $($ $p_{i}\equiv p=1/W_{\text{G}}$ for all $i)$ so that
the residual entropy is $S_{\text{R}}=\ln W_{\text{G}}$. The necessary (but
not sufficient) condition for this is that the system be macroscopically
large. The sufficient condition requires the system to be in internal
equilibrium, though not necessarily in equilibrium with the medium, as
discussed elsewhere.\cite{Guj-NE-I} In this case, the contribution $-\ln p=\ln
W_{\text{G}}$ to the entropy from any of the $W_{\text{G}}$ microstates is the
same so that%
\begin{equation}
S=-\ln p=\ln W_{\text{G}}; \label{Equiprobable_Probability}%
\end{equation}
there is no need to sum over all microstates since $\sum_{i}p_{i}=1$. Just one
microstate will give us the correct entropy. Similarly, just one microstate
will give a highly reliable value of any thermodynamic quantity. As the
equiprobable condition will be overwhelmingly satisfied for a system in
internal equilibrium, just one glass sample will give us a highly reliable
thermodynamics. The discussion can be applied to a system at any temperature,
provided it is in internal equilibrium. There is no need to carry out an
ensemble or time average. This is what makes classical thermodynamics such a
robust and highly reproducible endeavor, a result quite well known in
equilibrium. We have just extended it to systems in internal equilibrium.

Just because a glass sample at absolute zero is in a single microstate does
not mean that its statistical entropy is zero. Such a statistical entropy
interpretation does not agree with the thermodynamic entropy which, as
discussed above, is known to yield a non-zero residual entropy in many cases.
Therefore, such an interpretation, the one taken by Kivelson and
Reiss\cite{Kivelson} and by Gupta and Mauro,\cite{Gupta} must be considered
physically irrelevant. The correct statistical formulation of the entropy is
given by $-\ln p$, as shown in Eq. (\ref{Equiprobable_Probability}), and
merely reflects the sample \emph{probability} of a glass prepared under
identical macroscopic conditions.\cite{GujratiResidualentropy} Any sample will
give the same statistical entropy equal to the thermodynamic residual entropy
so that the statistical and thermodynamic interpretations of the entropy are equivalent.

The discussion should not imply that the ensemble average is no longer
meaningful just because we are dealing with a single sample. The way to see it
most clearly is to imagine dividing the system into a macroscopically large
number of quasi-independent subsystems, each of which itself is
macroscopically large. The subsystems are identical in size. Let $\iota$
denote a microstate of one of these subsystems, and $p_{\iota}>0$ its
probability. Then, it follows from their quasi-independence that $p=%
{\textstyle\prod\limits_{k}}
p_{\iota}$; the product is over all subsystems. It is now easy to see that
\[
S=%
{\textstyle\sum}
S_{k},
\]
where the sum is over all subsystems; each subsystem can be considered as
representing a member of the ensemble. All these members are considered at the
same instant.

Recently, Goldstein\cite{Goldstein-1} has also discussed the relevance of a
single microstate for the average at any temperature, but he does not discuss
the time required for the system to come to internal equilibrium or to
equilibrium after adding the enthalpy increment to it. Thus, it is by no means
clear that a single microstate during this time will really represent the
average properties of the system. His requirement of "...overwhelming majority
of microstates..." must refer to the equilibrium state of minimum Gibbs free
energy at given $T_{0},P_{0}$, but this is not relevant for a glass. As the
enthalpy increment is added, it takes a while for this enthalpy to distribute
itself throughout the system from the boundary. During this interval, the
system becomes \emph{inhomogeneous} and will not even be in internal
equilibrium, let alone in equilibrium. In this interval, we can still treat
each subsystem discussed above in internal equilibrium. But then we are
dealing with an ensemble average over inhomogeneous subsystems. We believe
that his conclusion would be valid only after the system has come into
internal equilibrium, though not necessarily in equilibrium, as demonstrated
above. For this, we must replace his above requirement by equiprobability of microstates.

\section{Conclusions}

The current work was motivated by the confusion about the residual entropy and
about the behavior of the entropy during relaxation that exists in the
literature, as discussed in Sect. \ref{Sect_Controversy_Status}. There are
opposing views not only about the residual entropy but also about the impact
of Goldstein's observation for the latter. The use of calorimetric data so far
has been to demonstrate that the irreversibility during a glass transition is
minimal so that $S_{\text{expt}}(T_{0})$ is not different from the actual
thermodynamic entropy $S(T_{0})$\ of the glass. But the calorimetric evidence
does not reveal how $S(T_{0})$ or $S_{\text{expt}}(T_{0})$ relate to
$S_{\text{SCL}}(T_{0})$. Therefore, how $S(T_{0})$ approaches $S_{\text{SCL}%
}(T_{0})$ remains unsettled; there are competing views in the literature. To
clarify the situation, we have considered the role of irreversible entropy
generation during isobaric vitrification and prove Theorems
\ref{Theorem_Lower_Bound} and \ref{Theorem_Entropy_Variation} that are valid
regardless of how far the system is out of equilibrium, as long as it is in
internal equilibrium; the latter is required to define the instantaneous
temperature, pressure, affinity, etc. The theorems are very general and are
not restricted by the "amount" of irreversibility. Theorem
\ref{Theorem_Lower_Bound} shows that the calorimetrically measured absolute
zero entropy forms a \emph{strict} lower bound to the residual entropy. The
former is shown to be positive under the assumption $S_{\text{SCL}}(0)=0$;
otherwise,\ we have the strict inequality in Eq. (\ref{Entropy_bound_at_0})
for a glass. In general, we have $S_{\text{expt}}(0)\geq S_{\text{SCL}}(0).$
It then follows that the residual entropy has to be larger than
$S_{\text{expt}}(0)$. Theorem \ref{Theorem_Entropy_Variation} shows that the
instantaneous entropy $S(T_{0},t)$ must always be higher than or at most equal
to the entropy $S_{\text{SCL}}(T_{0})$ of the equilibrated supercooled
entropy, which invalidates the inequality in Eq. (\ref{GM_bound}). During
isothermal relaxation, the entropy must decrease towards $S_{\text{SCL}}%
(T_{0})$ in time. All physical systems must follow the two inequalities in
Eqs. (\ref{ResidualEntropy_Bound}-\ref{GlassEntropy_Bound}) without any
\emph{exception}. We have demonstrated that the entropy loss proposal violates
the second law and have put the original observation of Goldstein on firmer
grounds. We have not only justified but also strengthened the calorimetric
evidence of the residual entropy by establishing $S_{\text{R}}>S_{\text{expt}%
}(0)>S_{\text{SCL}}(0)$.

The theorems follow from considering the thermodynamic entropy that appears in
the second law for an isolated system. Thus, any attempt to provide a
statistical version of entropy must satisfy these two consequences. We have
shown that the conventional statistical entropy formulation is consistent with
the thermodynamic notion of the residual entropy and that the equiprobability
requirement explains why a single sample is sufficient to give a highly
reliable thermodynamics of the system even when the latter is not be in
equilibrium with the medium.

\begin{acknowledgments}
I wish to acknowledge useful comments from Gyan P. Johari and Sergei V. Nemilov.
\end{acknowledgments}


\begin{thebibliography}{99}
\bibitem {Note-Measurement}The preparation to get the system in a unique
microstate requires knowing precisely the cell $C_{i}$ a particular particle
$i$ belongs to. This should be contrasted with the situation such as in a
crystal or a glass, in which, although each particle is confined within a
cell, we have no knowledge which particle belongs to this cell. This ambiguity
in the microstates is a characteristic of a statistical
system\cite{GujratiResidualentropy,Gujrati-Symmetry} that results in a
macrostate. The preparation of a unique microstate results in a unique
specification and is not physically feasible to accomplish in reality.
\bibitem {note1}By the second law we mean the law of increase of the entropy
according to which the (thermodynamic) entropy of an isolated system can never
decrease. Any statistical interpretation of this law must ensure this law.
\end{thebibliography}
\end{document}